\documentclass[twocolumn,showpacs]{revtex4}  %Printversion
\usepackage[latin1]{inputenc}
\usepackage{graphicx}
\usepackage{upgreek}
\usepackage{amsmath}
\usepackage{epsfig}
\usepackage{subfigure}
 \usepackage{placeins}
 \usepackage{amsmath}
 \usepackage{amssymb}
 \usepackage{wrapfig}
 \usepackage[]{tocbibind}
\usepackage{amssymb}
\usepackage{lineno}
 \usepackage{ifthen}
 \usepackage{times}

\makeatletter
\newcommand\figcaption{\def\@captype{figure}\caption}
\makeatother

\newlength{\mylinewidth}
\begin{document}



\title{The PANDA GEM-based TPC Prototype}

% if there is only one institution, use this form:
%\author{John Author, Giovanna Autore}
%\address{University of Wisdom, Physics City, Scienceland}

% else, use optional labels to link authors explicitly to addresses,
% as shown below:
\author{L. Fabbietti$^1$ for the GEM-TPC Collaboration}
\affiliation{$^1$ Excellence Cluster 'Universe', Technische Universit\"at M\"unchen}
\date{\today}
\begin{abstract}
We report on the development of a GEM-based TPC prototype for the PANDA experiment. 
The design and requirements of this device will be illustrated, with particular emphasis on the properties of the recently tested GEM-detector, the characterization of the read-out electronics and the development of the tracking software that allows to evaluate the GEM-TPC data.

\end{abstract}
\maketitle




\section{Introduction}
A GEM-based Time Projection Chamber (TPC) is one of the two options for the central tracker of the PANDA \cite{Panda} experiment at the new Facility for Antiproton and Ion Research (FAIR) at Darmstadt, Germany.
This facility will provide a cooled antiproton beam with momenta of 1.5-15 GeV/c, a maximal luminosity of 2$\cdot 10^{32}$ cm$^{-2}$ s$^{-1}$ that translates into 2 x 10$^7$ annihilations per second. The GEM-TPC tracker should provide a spatial resolution of  $\upsigma_{r\upphi}\approx$ 150 $\upmu$m, $\upsigma_z \approx $ 1 mm , a momentum resolution of about 1\% and at the same time minimize the amount of material in front of the electromagnetic calorimeter. In addition, particle identification at low momenta should be achieved by measuring the specific energy loss. The challenge is the continuous operation of this detector in the antiproton storage ring HESR. The usage of GEM foils \cite{Rop04} suppresses the backflow of ions into the drift volume, resulting in a controllable space charge accumulation.
We are currently building a prototype TPC, with a design very similar to the one planned for PANDA, but with smaller dimensions. It will be tested and employed in two experimental campaigns with the FOPI spectrometer at GSI and the CB-ELSA detector in Bonn. As a first step, a small GEM-TPC detector (test chamber) was built and characterized with cosmic mouns \cite{Wei07}. An average spatial resolution of 200 $\upmu$m  was achieved with rectangular pads (1$\times$ 6.2 mm$^2$) and with a read-out electronics based on the ALTRO chip \cite{Altro}. Later on, the pad geometry and the read-out have been modified to improve the signal quality. In order to characterize the test chamber with an external track definition in a particle beam, a new multi-purpose tracking system has been set up at the electron stretcher and accelerator ELSA, Bonn. Four GEM planes and four Silicon detectors have been installed on a mobile test bench bench to be used as a reference for tracking. The test chamber has been subject to first tests with $\approx$ 500 MeV the electron beam available at ELSA. First results are reported in \cite{Max09}. In this work we describe the GEM-TPC prototype. The paper is organized as follows: Section 2 describes the planned field cage, Section 3 the built and tested triple GEM detector. In Section 4 we report on the read-out electronics and in Section 5 its performance and the tracking procedure. In Section 6 we describe the data acquisition system realized for the prototype. 
\begin{figure}[hbt] 
\centering 
\includegraphics[width=0.40\textwidth,keepaspectratio]{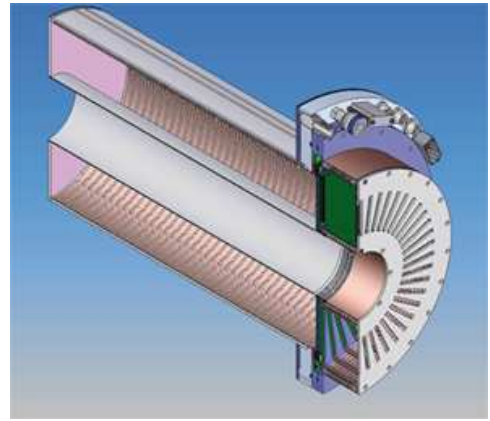}
\caption{3D Cross-section view of the GEM-TPC prototype with an outer (inner) diameter of 300 mm (105 mm) and a drift length of 730 mm.}
\label{scheme}
\end{figure}
\section{Field Cage and Media Flange}
A schematic view of the prototype is shown in Fig.\ref{scheme}. The cylindrical chamber has an active outer radius of 300 mm, an inner active radius of 105 mm and a 730 mm drift length. On the left side of Fig. \ref{scheme} the field cage is visible, connected to the GEM-detector via a media flange.\\
The field cage vessel is currently under construction and consists of a self-supporting sandwich structure made of 2 mm thick Rohacell core, Kapton insulation layers and two skins of fiber glass material, arranged in two concentric cylinders. The downstream end-cap is made of the same structure. Electrical shielding to the outside world is provided by an additional Kapton layer with aluminium coating.
 A homogeneous electric field in the drift direction is provided by the HV plane at the downstream end cap and precision concentric cylindrical field cage rings along the barrel that cover the inner and outer radius and are stepwise degrading the HV up to ground potential at the anode  side. To improve the homogeneity, the cylindrical field cage consists of two sets of copper strips on both sides of the Kapton foil. The potential on each ring is defined by a resistor chain.
The total thickness of the field cage has been estimated to be $ 1.2-1.8$ \% of a radiation length. \\
The upstream side of the field cage is connected to the media flange made of fiber glass material, which provides mechanical stability and serves as the mounting structure of the GEM-TPC to the external support. In addition the flange provides interfaces fo all external supplies like gas, cooling, high and low voltage. It hosts additional control devices which can be employed to monitor the temperature and the gas pressure inside the vessel.\\
A triple GEM detector, with the possibility of adding a fourth foil, of the type described in section \ref{GemDet} will be mounted to the upstream side of the media flange. 
The detector system is completed by the read-out plane equipped with the front-end cards described in section \ref{ele}. The TPC will be operated with a Ne/CO$_2$-based gas mixture (90/10), to keep the primary ionization small, reducing the space charge accumulation in the drift volume \cite{Boe09}. At the design value of the drift field of 400 V/cm, the maximum drift time in the Ne/CO$_2$ (90/10) gas mixture will be 27 $\upmu$s.
\section{The GEM Detector}
\label{GemDet}
\begin{figure}[hbt] 
\centering 
\includegraphics[width=0.40\textwidth,keepaspectratio]{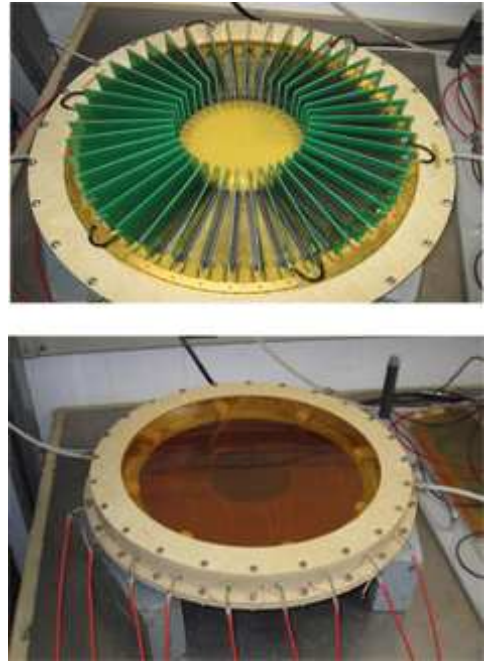}
\caption{Top and Bottom view of the triple-GEM detector. In the top panel the back side of the read-out plane, together with the 42 FE cards is shown. The bottom panels shows the entrance window side of the detector.}
\label{det}
\end{figure}

\begin{figure}[hbt] 
\centering 
\includegraphics[width=0.42\textwidth,keepaspectratio]{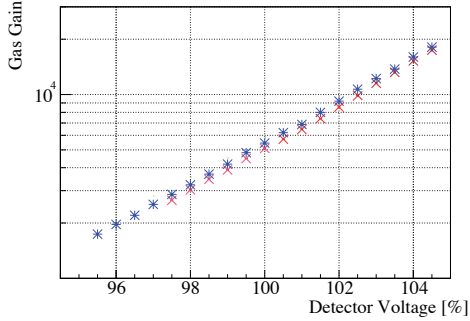}
\caption{Obtained gas gain as a function of the detector voltage. The stars refer to the current method, the crosses to the pulse height method. The data have been obtained with a Cu X-ray tube.}
\label{gain}
\end{figure}

\begin{figure}[hbt] 
\centering 
\includegraphics[width=0.42\textwidth,keepaspectratio]{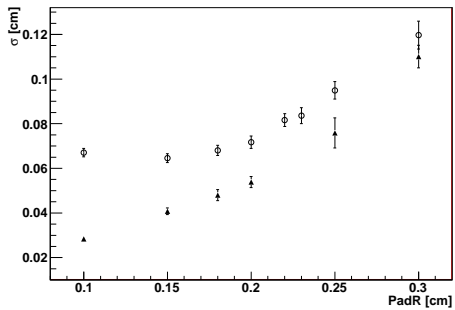}
\caption{Standard deviation of the residual distribution for reconstructed clusters as a function of the pad size. The results from simulations for the PANDA TPC (with a magnetic field of 2T) including and excluding diffusion effects (empty and full symbold respectively) are shown.}
\label{pad}
\end{figure}

A test detector with a triple GEM stack \cite{Xad09} has recently been built and tested, a picture of the device is shown in Fig. \ref{det}. It is composed of three circular GEM foils with a diameter of 300 mm interspaced with a 2 mm distance and a drift foil at a distance of 3 mm to the first GEM foil.
The GEM foil consists of a thin polyimide foil (50 $\upmu$m Kapton), metal-clad (5 $\upmu$m Cu) on both sides. It is perforated by a large number of holes (10$^4$/cm$^2$), with a diameter of 70 $\upmu$m and a pitch of 140 $\upmu$m arranged in a hexagonal pattern.
 Each GEM foil is divided on one side into 8 iris-shaped sectors that are powered independently to minimize discharges.
The detector is sealed from one side via a thin Kapton window (Fig.\ref{det} bottom panel) and on the other side with the read-out plane \cite{Berg09} (Fig. \ref{det} top panel), that consists of a four-layer printed circuit board (PCB) with 10296 pads.
A Ar/CO$_{2}$ mixture (70/30) has been used for the characterization of the detector.
The nominal settings of the high voltage, referred to in the following as the 100\% setting, translate to a drift field of 2.49 kV/cm and a transfer and collection fields of 3.73 kV/cm, respectively. The applied voltage to the three GEM foils yields 400 V, 365 V and 320 V from top to bottom respectively. First the operation of the detector has been verified by analyzing the response to a $^{55}$Fe source with the nominal settings.
In order to measure the gain of the triple GEM detector,  a Cu X-ray tube has been used employing two different methods.
The first method relies on the calibration of the preamplifier-amplifier-ADC chain by injection of different known charge values and measuring the corresponding ADC values.
The second method is based on the direct measurement of the current seen by the detector when an X-ray photon generates a known number of electrons in the gas detector. In this case the effective gain is obtained with the following formula:
\begin{equation}
G_{\mathbf{eff}}=\frac{I}{n_e \cdot e \cdot \phi},
\end{equation}
where $I$ is  the current measured at the readout anode, $n_e$ the number of ionization electrons per X-ray photon absorbed in the gas and $\phi$ the total X-ray rate. A precise current-meter has been developed and constructed for this purpose which is able to measure currents ranging from 10 pA to 10 mA in four switchable ranges. In order to be able to insert the device in high voltage lines up to 6 KV, a wireless readout via the XBee protocol has been implemented. The operation of more than 10 devices at the same time is possible, which enables the current measurement on multiple electrodes. The resulting gain curves are shown in Fig. \ref{gain}, where the gas gain is displayed as a function of the detector voltage, 100\% corresponding to the settings mentioned above. For the other settings, all voltages and fields were scaled according to the values given. One can see that the two methods deliver similar results with the current measurement  being a more direct measurement and independent of the accuracy of the ADC calibration.
The triple GEM detector is read out via a PCB hosting 10.000 hexagonal pads of 1.5 mm outer radius \cite{BerDip}. The pad geometry has been chosen to ensure a homogeneous charge distribution between neighboring pads and the optimal pad size has been estimated with simulations. Fig. \ref{pad} shows the cluster resolution achieved for pion tracks of 0.5 GeV/c momentum in the PANDA GEM-TPC as a function of outer pad radius assuming hexagonal pads. The two curves correspond to the results achieved with and without diffusion effects in the simulation. One can see that the effect of the electron diffusion 
saturates the resolution at a pad radius of about 1.5 mm. To validate this result a new pad-plane with pads of two different sizes, 1.5 and 1.25 mm, has been attached to the small test chamber \cite{Berg09,Max09}. Muon tracks have been collected  and the resolution evaluation is currently ongoing. New measurements with the electron beam at ELSA are planned.
\section{Read-Out Electronics}
\label{ele}
An analog sampling ASIC designed for the readout of micropattern gas detectors, the AFTER chip \cite{t2k} is used for the read-out of the GEM-TPC prototype. Each of the 72 channels of the AFTER chip includes a low noise charge preamplifier and a 511-cell capacitor array . It can be operated at sampling rates between 10 and 50 MHz and offers a large flexibility in the shaping time (100 ns- 2 $\upmu$s). 
The multiplexed data stream is digitized by a custom-made pipelined ADC module. \\
\begin{figure}[hbt] 
\centering 
\includegraphics[width=0.22\textwidth,keepaspectratio]{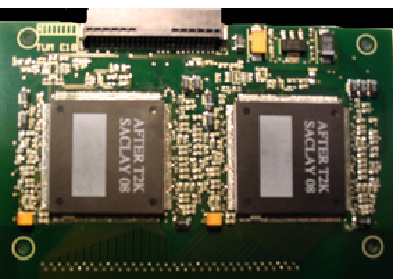}
\includegraphics[width=0.35\textwidth,keepaspectratio]{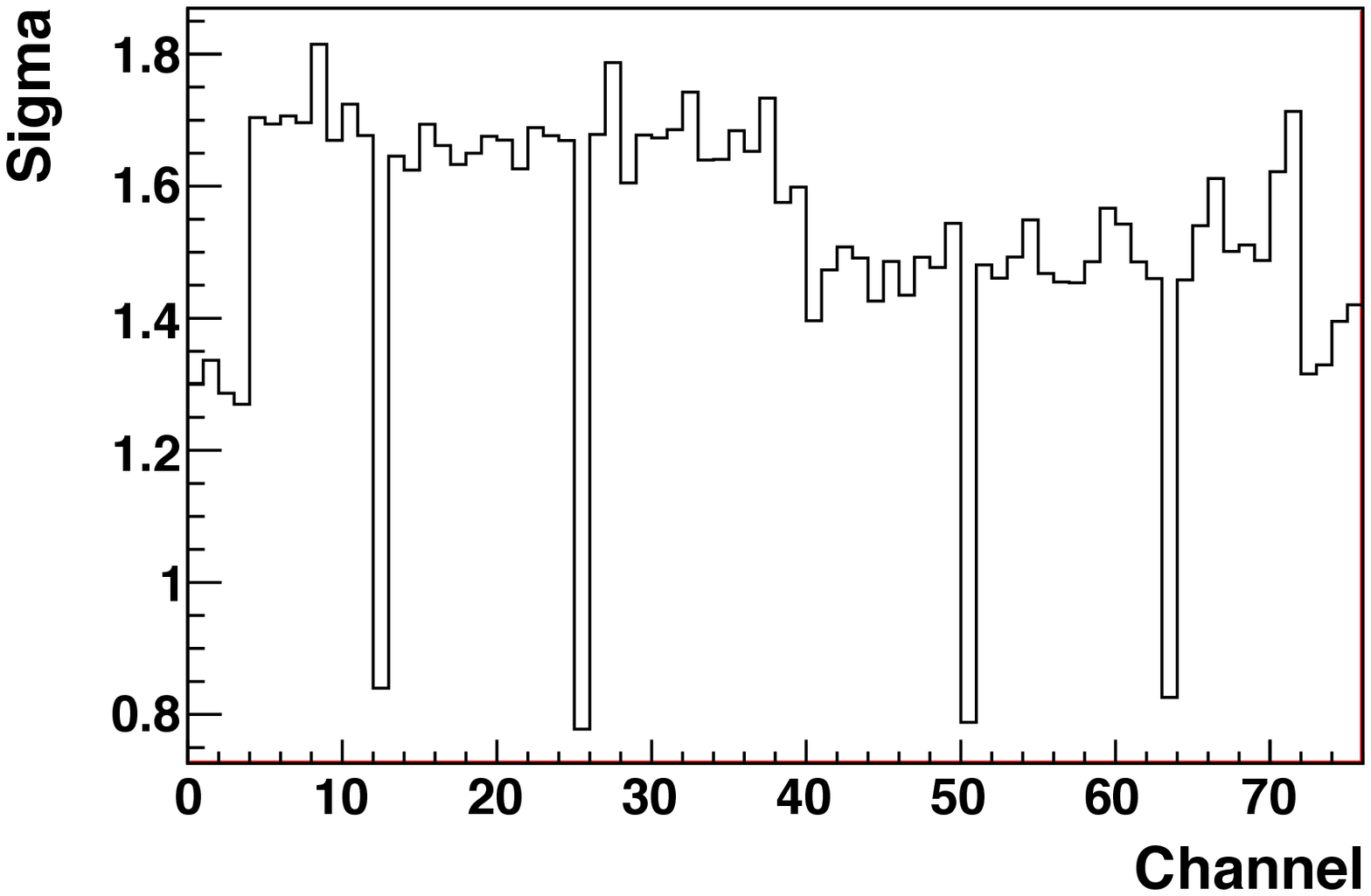}
\caption{(Top) Picture of the FE card with two of the four AFTER chips. (Bottom) Noise distribution in ADC channels of the 76 chip channels.}
\label{chip}
\end{figure}
42 rectangular front-end (FE) cards (see Fig. \ref{chip}, top panel), each equipped with 4 chips, have been produced and will be connected to the read-out padplane described in \cite{BerDip} to read out a total of 10.000 hexagonal pads.
The produced FE cards connected to the pad-plane have been tested in terms of noise and gain performance. Fig. \ref{chip} (bottom panel) shows the noise in units of ADC channels measured for one chip. Of the 76 channels of each chip, only 64 are connected to pads. Four channels on each side remain unused and thus have a lower noise. The step around channel 38 is due to slightly longer signal lines of the FE card for the first group of channels. The four unconnected channels with a very low noise of less than 1 ADC ch. are intended for common mode noise measurement and correction as well as to correct possible baseline shifts along the analog memory.
The sensitivity of the readout chain was determined by injecting voltage pulses of different amplitude through a 1pF capacitor, resulting in 0.055 fC/ADC ch.. Therefore, the average noise of all channels connected to pad is 530 e$^-$. The cross-talk between neighboring strips on the FE card has been measured to be around 1\%. 
Currently the AFTER cards are operated with a frequency of 20 MHz and a shaping time of 200 ns.
\section{Track Reconstruction}
\begin{figure}[hbt] 
\centering 
\includegraphics[width=0.37\textwidth,keepaspectratio]{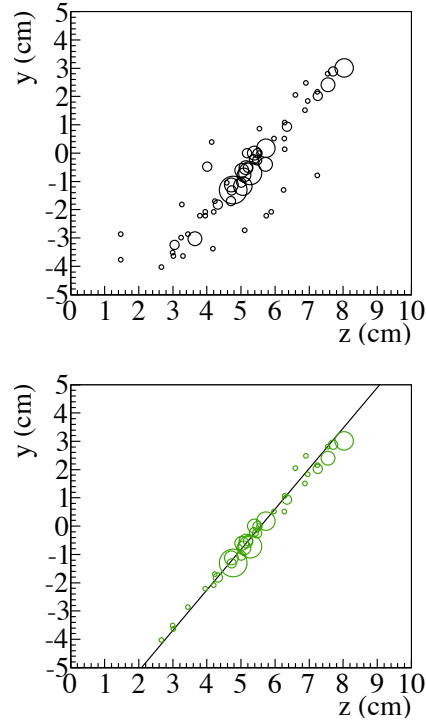}
\caption{(Top) Reconstructed clusters produced by a cosmic track in the test chamber. (Bottom) Cluster hits after the pattern recognitions steps together with the result of the Hough transform (solid line).}
\label{cos}
\end{figure}

In order to test the track reconstruction algorithm, the AFTER read-out has been connected to the test chamber described in \cite{Wei07,Berg09} and tracks produced by cosmic radiation have been recorded.
 The full track reconstruction chain, embedded in the PandaRoot framework, has been developed and first tested with simulated data. Full scale simulations have been employed to test the clusterisation, pattern recognition via fast Hough transformation and  track fitting using the GENFIT tool \cite{GenF}. This same analysis procedure has been applied to the experimental data. Fig. \ref{cos} shows an example of a cosmic track detected in the test chamber. In the upper panel the y vs z position of the reconstructed clusters are shown, the radius of the circles is proportional to the signal amplitude. In the lower panel the cluster signals are shown after the pattern recognition steps, together with the results of the fast Hough transform (solid line). One can see how noise hits are suppressed by the pattern recognition and how the Hough transform result matches well with the trajectory.
 All the tools for the alignment of the test chamber with the other detectors of the ELSA telescope have been developed. A coarse measurement of the relative positions of the detectors has been done using photogrammetry, while the precise alignment of the telescope is computed via the Millipede program \cite{Milli} using electron tracks. The TPC is then aligned with respect to the telescope using a Minuit-based $\upchi^2$ minimization. Lately, the entrance window of the test chamber was replaced to minimize the material along the electron tracks. Indeed so far multiple scattering effects made it impossible to evaluate the intrinsic resolution of the TPC equipped with the new read-out. Data collected at a higher electron beam energy at ELSA or using a muon beam at CERN are expected to deliver better results in the near future.
\section{Integration of the GEM-TPC in FOPI}
The GEM-TPC prototype will be installed in the FOPI apparatus at GSI in order to test its performance in a running spectrometer, featuring a 0.8 T solenoid field, tracking detectors (Central Drift Chamber) and good particle identification using RPC \cite{RPC}. 
Therefore the TPC data acquisition
system has to be integrated in the Multi Branch System (MBS) used for the FOPI data acquisition.  
The trigger signals from FOPI are used as input for the TPC Trigger Control System (TCS), an optical system which distributes trigger information as well as a common clock to the TPC readout adopted from the COMPASS experiment \cite{Com}. This system also generates the dead time required for the readout of the TPC, which is then merged with the dead time required for the FOPI detectors to inhibit triggers too close in time to an accepted one. Since the FOPI data are read out via the VME backplane which would be too slow for the large amount of data from the TPC, a standalone readout using the S-Link protocol is used for the TPC \cite{Com}. Both FOPI and TPC data streams are merged at the event-building stage using unique event numbers or time stamps which have been distributed to both data streams via the TCS system. To this end a sender-receiver architecture, based on the TCP protocol, has been set up which allows to include the TPC data in the MBS.\\
This acquisition system will be used to collect the data of the GEM-TPC during the test within the FOPI spectrometer. The TPC will be inserted in the Central Drift Counter (CDC) cylindric tracking chamber of FOPI, surrounded by  time of flight detectors. We plan to exploit the Li beam with  2 AGeV and a maximum intensity of 10$^8$ particle/sec  on a polyethylene target. Tracks reconstructed in the CDC serve as reference for the GEM-TPC track reconstruction. Considering a target with an interaction length of 1\% a maximal reaction rate of about 10 KHz is expected. The effect of the FOPI first level trigger, based on the measurement of charged particle multiplicity, will reduce the event rate to 1 kHz, which is compatible with both the FOPI and GEM-TPC readout. The operating GEM-TPC detector should allow to extend significantly the acceptance for the reconstruction of $\Lambda$ and K$^0_S$ and might be very useful to investigate both in-medium hadron properties as the study of kaonic bound states in proton and pion-induced reactions.
%\label{}

\end{document}